# MÖSSBAUER EXPERIMENT IN A ROTATING SYSTEM ON THE TIME DILATION EFFECT

Alexander L. Kholmetskii[1], Tolga Yarman[2], Oleg V. Missevitch[3] and Boris I. Rogozev[4]

[1]Department of Physics, Belarus State University, Minsk, Belarus, kholm@bsu.by
[2]Okan University, Istanbul, Turkey & Savronik, Eskisehir, Turkey
[3]Institute for Nuclear Problems, Minsk, Belarus
[4]V.G. Khlopin Radium Institute, S.-Petersburg, Russia

## Abstract

We present results of Mössbauer experiment in a rotating system, which is induced by our recent disclosure (Phys. Scr., **77** (2008) 035302) and which consisted in the fact that a correct processing of Kündig's experiment data on the subject gives an appreciable deviation of a relative energy shift $\Delta E/E$ between emission and absorption resonant lines from the standard prediction based on the relativistic dilation of time (that is $\frac{\Delta E}{E}=-\frac{v^2}{2c^2}$ to the accuracy $c^{-2}$, where $v$ is the tangential velocity of absorber of resonant radiation, and $c$ is the light velocity in vacuum). Namely, the Kündig result following the correction we brought to it, is $\frac{\Delta E}{E}=-(0.596\pm 0.006)\frac{v^2}{c^2}$. In our own experiment we carried out measurements for two absorbers with substantially different isomer shift, which allowed us to make a correction of Mössbauer data to a level of vibrations in the rotor system at various rotational frequencies. As a result we got the estimation $\frac{\Delta E}{E}=-(0.68\pm 0.03)\frac{v^2}{c^2}$. A deviation from the relativistic formula is discussed.

## 1. Introduction

Soon after the discovery of the Mössbauer effect, at the early sixties, a series of experiments with resonant γ-quanta in rotating systems were carried out [1-6]. All these experiments used a typical configuration, where an absorber orbited around a source of resonant radiation (or vise versa), while a detector was placed outside the rotor system in the plane of rotation. A common goal of these experiments was to verify the relativistic dilation of time (or, which is the same for this configuration, the Second Order Doppler shift, SOD) for a moving resonant absorber/ source. A relative energy resolution of resonant γ-quanta has a typical value of $10^{-13}$ to $10^{-14}$, which allows a reliable measurement of the time dilation effect for sound (or even sub-sound) tangential velocities. Among these experiments, a separate attention should be accorded to the experiment by Kündig [1], because he was the only one who successfully applied a first order Doppler modulation of energy of γ-quanta on a rotor at each fixed rotation frequency, implementing an oscillating motion of the source along the radius of the rotor. By such a way he measured the shape and the position of resonant line on the energy scale versus the rotation frequency. In contrast, all other authors [2-6] measured only the count-rate of detected γ-quanta as the function of rotation frequency. Thus, it is needless to say that Kündig's experiment is much more informative and reliable than other experiments on the subject. In particular, he observed an essential (up to 1.5 times) broadening of the resonance line with increase of rotational frequency, caused by mechanical vibrations in the rotor system. In Kündig's experiment this line broadening does not lead to additional systematic errors, but such a privilege does not hold in the other experiments mentioned above. It seems that in the scale of Mössbauer measurements, such vibrations are practically unavoidable, at least at the range of rotational frequencies, corresponding to non-negligible elastic deformations of rotor.

Nonetheless, the authors of [2-6], just like Kündig [1], reported the confirmation of the relativistic expression for the time dilation effect, written to the accuracy $c^{-2}$

$$\frac{\Delta E}{E} = -\frac{v^2}{2c^2} \qquad (1)$$

with the measuring precision about 1 %.

However, the first three authors of this paper revealed some time ago [7] that data processing in Kündig's experiment was erroneous (errors were not though related to vibration). An intriguing fact is that after correction of the errors by Kündig, his experimental data gave the value [7]

$$\frac{\Delta E}{E} = -k\frac{v^2}{c^2}, \qquad (2)$$

where $k = 0.596 \pm 0.006$ (instead of 0.5). One sees that the deviation of eq. (2) from the relativistic expression (1) exceeds almost 20 times the measuring error. What is more, we have shown in [7] that the experiment [6], containing much more data than [2-5], is also well fitted into the supposition $k$>0.5.

One may regret that the finding by Kündig (2) was masked during almost 50 years by the data processing errors he may have committed. Thus, it seemed topical to repeat once again the Mössbauer experiments in rotating systems to verify the validity of either eq. (2), or eq. (1). At the same time, a modulation of energy of γ-quanta in rotating system, realized by Kündig, remains for today a non-trivial problem.

In the present paper we suggest a method for correction of Mössbauer data against possible vibrations between a radioactive source and resonant absorber in rotating systems without the first order Doppler modulation of the energy of gamma-quanta (section 2). In section 3 we describe the experimental setup developed by ourselves for the measurement of Mössbauer effect in rotating system. In section 4 we present the collected experimental data and results of their processing. Finally, section 5 contains an open discussion of the results obtained.

## 2. Methodological approach

Our goal is to provide a reliable measurement of a relative energy shift $\Delta E/E$ between the resonant lines of Mössbauer source and rotating absorber in a wide range of rotating frequencies, when the first order Doppler modulation of energy of resonant γ-quanta on the rotor is not applied.

In what follows, we consider $k$ in eq. (2) as a parameter to be determined experimentally. In an idealized case, where the mechanical vibrations are absent in rotating system, the coefficient $k$ in eq. (2) is easily computed, if we compare the second order Doppler shift, obtained as the function of rotation frequency, with the first order Doppler shift to be measured for the same pair of "source plus absorber" with a standard Mössbauer spectrometer outside the rotor system.

However, the analysis becomes more complicated in a real case, when the non-vanished vibrations are present in rotating system. Such vibrations are known to broaden the given resonant line, but do not influence the total area and position of this resonant line, on the energy scale (Fig. 1). Further on, if we could assume that the level of vibrations remains constant with variation of rotational frequency, the problem of determining the coefficient $k$ in eq. (2) is again easily solved. In this case we choose a pair of "source plus absorber" with the isomer shift to be equal approximately to the half of the full range of variation of energy due to SOD (Fig. 2). As we mentioned above, the vibrations do not change the position of resonant line upon the energy scale. Hence, in a rotor experiment it is enough to find a tangential velocity $v_m$, which replies to a minimal count-rate of detector (i.e., the extreme of resonant line). Since the corresponding energy shift for this extreme $(\Delta E/E)_m$ is exactly known from the available Mössbauer spectrum, we directly compute the coefficient $k = \frac{c^2}{v_m^2}(\Delta E/E)_m$.

At the same time, in a general case we have to adopt that the vibrations are not only present in the system, but also vibration level depends on the rotation frequency. Just such a behavior of vibrations has been found by Kündig, and for convenience we reproduce in Fig. 3 the Mössbauer spectra, obtained in Kündig's experiment at various rotation frequencies. In this case the implementation of the measurements with the absorber of Fig. 1 only (named hereinafter the absorber 1), occurs insufficient for the exact determination of $k$. One can show that the extreme of count-rate of detector as a function of rotation frequency $\nu$ is shifted either at lower frequencies (when vibrations increase with $\nu$), or at higher frequencies (when vibrations decrease with $\nu$) in comparison with the realistic position of the resonant line on the energy scale, and accordingly, one may end up with distorted information on the value of $k$ in eq. (2).

Our next idea is to eliminate such distortion of $k$ due to vibrations, applying additionally another absorber "2" with a Mössbauer line, whose shape drastically differs from the shape of the line for absorber 1, at least on the range of variation of energy due to SOD. Correspondingly, the broadening of resonant lines due to vibrations should induce the essentially different variations of detector's count-rate for each absorber. Thus, comparing the data of the rotor experiment with absorbers 1 and 2, we can get full information on the level of vibrations and make the required corrections under evaluation of the value of $k$.

In practice, it is enough to deal with a resonant line of the second absorber with an appropriately larger isomer shift than for the first absorber, so that a full range of variation of energy of $\gamma$-quanta due to SOD lies on a slope of the line, Fig. 4. Now let us show that a conjoined processing of data obtained in a rotor experiment with the absorbers 1 and 2 allows us actually to eliminate the influence of vibrations on the estimation of $k$. For this purpose we realize an algorithm as follows.

1. The Mössbauer spectra of both absorbers are measured with a high statistic quality.
2. The data of rotor experiment are collected for absorbers 1 and 2 at the same range of variation of rotation frequency.
3. Having assumed any particular value of $k$ in eq. (2) (for example, one can vary $k$ between 0.5 and 1.0), the expected theoretical curves for a rotor experiment with a zero level of vibrations (hereinafter the "idealized rotor experiment") are plotted for both absorbers. Since in a real experiment the vibrations are expected to be anyway present, the theoretical curves we draw for each absorber, deviate from corresponding experimental curves, no matter what the assumption on $k$ is.
4. As we mentioned above, one can propose to model the influence of vibrations by varying the width $\Gamma$ of resonant line at various $\nu$, fixing its position and total area to be constant. A variation of $\Gamma$ for the resonant line of absorber 2 is implemented for the chosen $k$ at each rotation frequency $\nu$, so that the corrected theoretical curve passes exactly through the available experimental points.
5. The dependence $\Gamma(\nu)$ obtained with the absorber 2 at the given $k$ is applied to correct the theoretical curve for the absorber 1, obtained for the same $k$. If this new curve continues to deviate from a set of corresponding experimental points, then we adopt that the hypothesis on a given value of $k$ is false and should be rejected.
6. A new value of $k$ is chosen, and the steps 3-5 are installed again, while we obtain a self-consistent result for the absorber 1 with the minimal statistical test criterium $\chi^2$. A corresponding magnitude of $k$ is then adopted.

We would like to add that a modern extended database on various Mössbauer compounds created by the Mössbauer Effect Data Center [8] allows us choosing the most optimal absorbers for realization of the algorithm just described. In section 4 we use this algorithm to processing of data obtained, applying the MathCad Professional software.

## 3. Experimental

A rotor system developed by ourselves is based on the ultracentrifuge K-80 (Belmashpribor, Minsk) with the diameter of working chamber 630 mm and the range of rotation frequencies $\nu$=0 to150 rev/s. The rotor has the form of flat streamlined rod, and it was made from a special ultrastrong and light aluminum alloy doped by titanium and exposed to special thermal treatment. The diameter of the rod is 610 mm. A Mössbauer source $^{57}$Co(Cr) with the activity 20 μCi was put into a Cu-Pb shielding & collimating system and mounted on the rotation axis. A sample holder was made from a special tempered aluminum alloy and fixed at the edge of the rotor. A semi-hermetic chamber of the rotor system was continuously pumped out during measurements; a stationary pressure was about 100 mmHg. In these conditions there was some heating of the rotor during its rotation. However, the difference of temperatures of the source and absorber never exceeded 10°C within the applied range of rotation frequencies. For such a difference of temperature, we can neglect by variation of resonant line position.

A proportional counter for detection of γ-quanta, filled by xenon, was located outside the rotor system. A beryllium window ∅15 mm for output of resonant radiation was made in the working chamber opposite the working window of detector, and its center belonged to the same plane as the line, joining the source and absorber. For the reasons, explained below (see, section 5), we chosen the distance between the source and absorber ($r$=305 mm) many times larger than picked for the Kündig experiment (where $r$=46 mm). Then the maximum tangential velocity of the rotor is $v=2\pi\nu r=2\pi\cdot 150\cdot 0.305=287.3$ m/s.

In order to compensate partially the decrease of count-rate of detector, caused by the enlarged distance between source and absorber, we applied the absorbers elongated in the azimuthal direction. It allows increasing as much as possible the effective measuring time per each rotation period (see, Fig. 5). The absorbers used have a rectangular form with the size 15×55 mm. In these conditions the average count-rate of detector was about 3 pulses/s in the working energy window, selecting resonant $\gamma$-quanta of $^{57}$Fe 14.4 keV. Due to a careful shielding of the source, a background count-rate (measured at the angular positions of rotor, when radiation of the source does not pass across detector's window) was less than 0.2 pulses/s. This allowed us to refuse from time selection of output pulses of detector, applied in the earlier experiments [1-6].

A measurement of SOD was carried out at the range of rotation frequencies $\nu$=70 to 120 rev/s. It corresponds to the change of tangential velocities of absorber $v$=134.1 to 230.0 m/s. In terms of the first order Doppler shift, a range of variation of linear velocities is

$$u = kc\frac{v^2}{c^2} = k(0.060 \text{ to } 0.176) \text{ mm/s}.$$

Each measuring cycle started with the maximal frequency $\nu$=120 rev/s with its further decrease to 70 rev/s with steps of 10 rev/s. The accuracy of setting of $\nu$ is ±0.5 rev/s. The number of output pulses of the detector was measured during 100 seconds at each rotational frequency. Then a new measurement cycle was installed, and so on. A total number of counts at each $\nu$ has been obtained by summation over 50 cycles.

The absorber 1 represents a thin layer of the compound $K_4Fe(CN)_6\times 3H_2O$ enriched by $^{57}$Fe to 90 %. The absorber 2 is a thin layer of the compound $Li_3Fe_2(PO_4)_3$ enriched by $^{57}$Fe to 90 %. Each absorber was placed into beryllium shielding transparent for resonant $\gamma$-quanta 14.4 keV. The Mössbauer spectra of absorbers were measured by means of the Mössbauer instrument package MS-200IP [9].

**4. Results**

In Fig. 6 we show the Mössbauer spectra of the absorbers 1 (a) and 2 (b). The spectrum of the first absorber represents a single line shifted at +(0.095±0.001) mm/s with respect to emission line of $^{57}$Co(Cr). The value of resonant effect is (20.9±0.1) %. The spectrum of the second absorber represents a partially resolved doublet with left line shifted at +(0.390±0.001) mm/s with respect to $^{57}$Co(Cr). The resonant effect is (30.5±0.1) %. The expanded relevant fragments of both spectra are depicted in Fig. 7, where the dot vertical lines restrict a range of variation of SOD for $\nu$=70 to120 rev/s and $k$=0.5, whereas the continuous vertical lines show the same range for $k$=1.0.

One can see that the parameters of Mössbauer spectrum of absorber 1 make it especially sensitive to the choice of two limited cases $k$=0.5 and $k$=1.0, if one considers the idealized rotor experiment (exempt of vibrations). If $k$=0.5, then a count-rate of detector should continuously decrease in a full range of variation of rotation frequency $\nu$, and at $\nu$=120 rev/s we only approach to the minimum of resonant absorption. If $k$=1.0, then the count-rate of detector reaches a minimal value already at $\nu\approx$90 rev/s with further increase at higher $\nu$.

A similar analysis of the idealized rotor experiment with absorber 2 indicates that a count-rate of detector decreases for both limited cases ($k$=0.5; $k$=1.0), but a slope of falling curve is few times larger at $k$=1.0.

In Figs. 8, 9 we present the data of our rotor experiment, obtained with both absorbers. In these figures we also plotted the corresponding curves, expected in the idealized rotor experiment at various $k$. According to our approach to the data processing described above (section 2), we use the data of Fig. 9 (absorber 2) to evaluate the level of vibrations in the rotor system at different $\nu$. Then we re-compute the expected curves for the absorber 1 at various $k$ with account of vibrations in the system and compare them with the experimental data of Fig. 8.

The analysis of Fig. 9 (absorber 2) indicates that vibrations are actually present in the system (like they were in the Kündig experiment) and do distort the count-rate of detector in comparison with the idealized case. In particular, the number of counts obtained at $\nu$=120 rev/s does not comply with the computed curves in a full adopted range of variation of $k$=0.5 to1.0. Furthermore, since these computed curves have a different slope for different $k$, a level of vibrations expressed through the dependence $\Gamma(\nu)$, occurs sensitive to the choice of a particular $k$. At the same time, at this stage we can reduce the volume of computing work rejecting the two limited hypotheses $k$=0.5 and $k$=1.0 as physically non-adequate, assuming that just one $k$ is to be adopted throughout. Indeed, if we suppose the validity of $k$=0.5 (this is just the classical relativistic prediction) then we conclude from the data of Fig. 9 (the experiment with absorber 2) that up to $\nu$=110 rev/s, the vibration in the rotor system are negligible. In such a case the experimental points depicted in Fig. 8 (the experiment with absorber 1) must lie on the curve computed for $k$=0.5 at least at the range $\nu$=70 to110 rev/s. However, we observe a drastic deviation of the experimental data from this curve. Further on, if we adopt that the data of experiment with absorber 1 (Fig. 8) match well with the computed curve for $k$=1.0, we must conclude that the level of vibration is negligible in a full range of variation of $\nu$. However, a comparison of the computed curve for the absorber 2 at $k$=1.0 with experimental data (Fig. 9) shows that vibrations in the rotor system should be relatively high for $\nu\geq$90 rev/s. Hence we again get a self-contradictory result. Thus, we restricted our further analysis by the range $k$=0.6 to 0.9.

In Fig. 10 we again present the results of the rotor experiment with absorber 1 in comparison with the expected theoretical curves re-computed for $k$=0.6, 0.7 and 0.8 via taking into account of the dependence $\Gamma(\nu)$, obtained from the data processing with absorber 2. It is seen visibly that the best correspondence between the experimental data and theoretical curve occurs nearly $k$=0.7. A least square fit confirms this observation and gives $k$=(0.68±0.03).

**5. Discussion**

Like in the corrected processing of data of Kündig's experiment, we have the privilege of having made the contribution to [7], once again we reveal an appreciable deviation of the coefficient $k$ in eq. (2) from the classical relativistic prediction $k$=0.5. Of course, we trust in the validity of the usual relativistic dilation of time due to the motion, which has numerous confirmations in the experiments dealing with atomic beams and free muons (see, *e.g.*, [10, 11]). Rather, we conjecture that with reference to the Mössbauer experiments on a rotor, the energy shift of absorption resonant line is induced not only via the standard time dilation, but also via some additional effect missed to the moment. Discussing in [7] a possible origin of this effect, we supposed that the pressure created by the centrifugal force on the absorber, might change the electron density on the resonant nucleus and, by such a way, change the isomer shift between emission and absorption lines. This leads to the increase of measured $k$. Although this effect was not supported by numerical estimations of [7], we decided to eliminate any speculations on the influence of pressure, choosing the rotor diameter 6-7 times larger than in Kündig's and other Mössbauer experiments in rotating systems. In this case the centrifugal pressure on the absorber is the same times smaller for a fixed tangential velocity $v$. Nonetheless, we observe even a higher value of $k$ than drawn by the Kündig experiment. This provides us with reason to assume that the missed effect, inducing the increase of $k$ in comparison with the standard value 0.5, has deep physical roots. In this connection it deserves to pay a careful attention to the research of the second co-author of this paper on the properties of quantum particles and macroscopic objects in bound states [12-14]. In particular, Yarman has shown that a change of binding energy of particles can be described in a non-contradictory way through a corresponding change of their effective rest masses. Furthermore, exploring the properties of symmetry of Schrödinger equation, in general any appropriate relativistic or non-relativistic quantum mechanical description in relation to the object at hand, one can show [12-14] that such a change of the effective rest mass of bound particle is accompanied by a corresponding change $\delta t$ of its time rate. At a qualitative level, this effect is expressed as

$$\frac{\delta t}{t} = \frac{E_b}{mc^2}, \tag{3}$$

where $E_b$ is the biding energy, and $m$ is the rest mass of the particle. This equation does not contain any quantum-mechanical constant and hence it can be naturally extended to the macro-world. On this basis Yarman has obtained (up to a third order Taylor expansion) the end results of general relativity [12].

For the Mössbauer experiment in rotating system, the absorber represents, in fact, a macroscopic object in a mechanically bound state and thus, besides of the relativistic dilation of time, one can expect an additional decrease of time rate due to eq. (3). An actual binding energy of a rotating absorber to the centrifugal field depends on a rotor material, and might well induce as Yarman et al. predicted it [14], a dilation of time of the same order of magnitude, as that of the classical relativistic effect due to a motion. Hence eq. (3) might be considered as a vital effect for explanation of the observed value $k$>0.5. In addition, this could explain a deviation of our result ($k$=0.68±0.03) from the result derived from Kündig's measurements ($k$=0.596±0.006), insofar as in both experiments the rotors of different materials and diameters were applied.

At this stage, we do not wish to insist that the prediction by Yarman et al (which though originally had been the driving force of the present work) indicates a single possible way to explain the deviation of the coefficient $k$ from 0.5. In any case, at this stage, the implementation of new Mössbauer experiments in rotating systems seems to be an essential task, in order to collect new data on the relative energy shift between emission and absorption lines in various conditions. Planning such new experiments, one should recognize that the approach by Kündig (the first order Doppler modulation of resonant radiation on a rotor) remains the best; in particular, one sees that the measuring error of $k$ in his experiment is few times lower than in the reported experiment. This is not only due to a better statistic quality of Kündig's data obtained for the higher count-rate of detector, but also due to elimination of any correction of these data to a level of vibrations in a rotor system. At the same time, we hope that the present experiment, amongst other things, has its own independent significance as a definite confirmation of the corrected result of Kündig's experiment $k$>0.5.

**Acknowledgements**

It is our pleasure to thank the Okan University (President of The Board of Trustees Bekir Okan, Rector Prof. Sadik Kirbas, Istanbul, Turkey), and Savronik (Defense Electronics) Company (President Mustafa Kula, and Vice President Kenan Isik, Ancient Students of Prof. T. Yarman, Eskisehir, Turkey) for the financial support, they have kindly provided (Contract No 08927 with the Belarus State University). We thank Prof. Metin Arik for many hours of precious discussions. Thanks are further due to Dr. S. Yarman, and Dr. F. Yarman Ex Presidents of Savronik, for having honorably backed up the endeavor in question. We thank Eng. J. Sobolev, Eng. V. Furs (Belmashpribor, Minsk, Belarus) and Eng. A. Zhenevskii (retired, Minsk, Belarus) for installation of the rotor system. We thank Dr. M. Silin and V. Soloviov (V.G. Khlopin Radium Institute, S.-Petersburg, Russia) for preparation of both absorbers. Finally we thank Dr. A. Khasanov and Prof. J. Stevens (Mössbauer Effect Data Center, Asheville, USA) for the extensive consultations regarding the choice of resonant absorbers for our research.

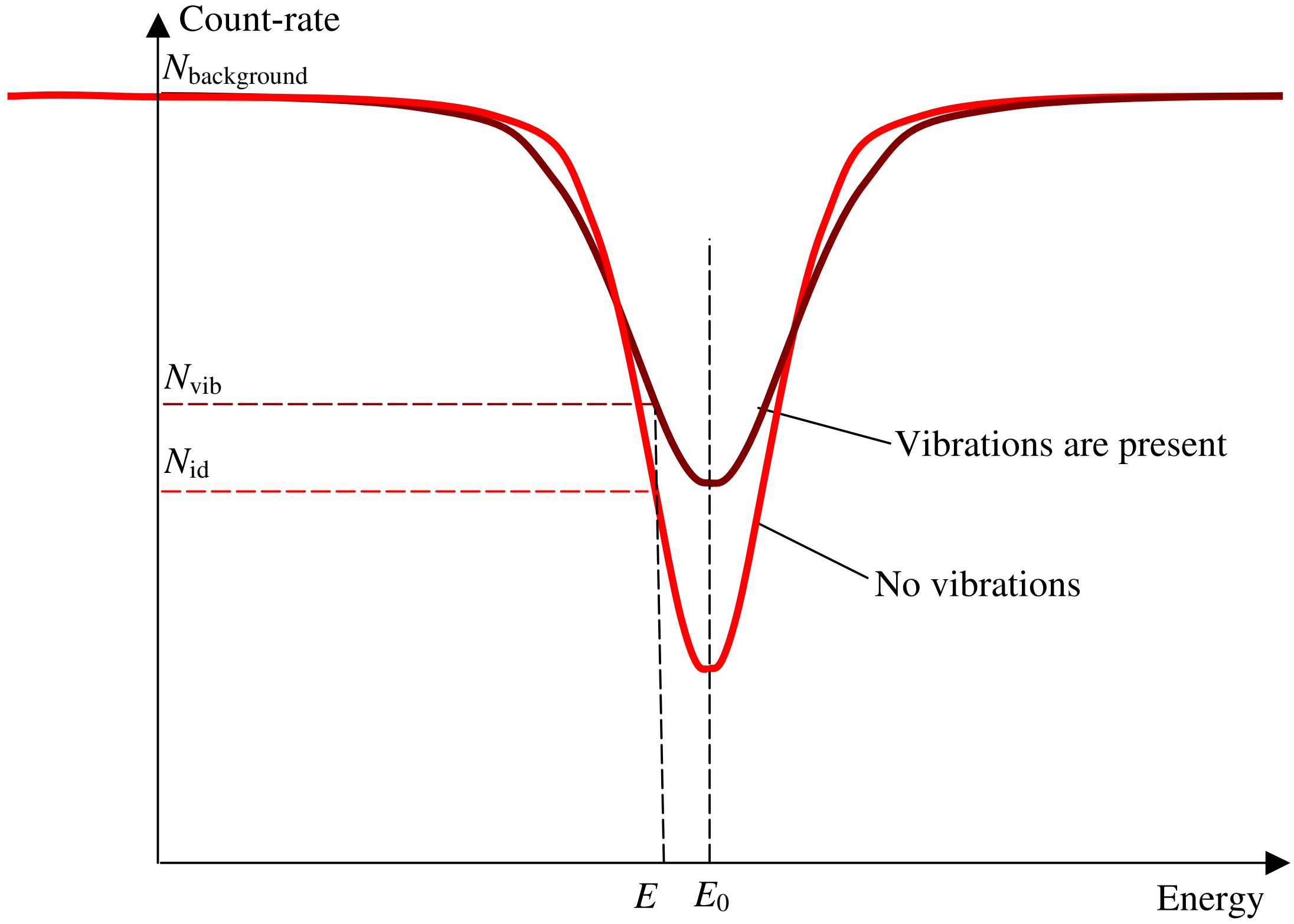

Fig. 1. Broadening of a resonant line due to vibrations. $E_0$ is the position of the line on the energy scale. One sees that that the count-rate of a detecting system at some energy $E$ in the presence of vibrations ($N_{vib}$) differs, in general, from the count-rate for the idealized case ($N_{id}$). If the energy $E$ lies not far from the extreme $E_0$, then $N_{vib} > N_{id}$. It is assumed that the vibrations do not change the Lorentzian form of resonant line, which is correct, when a range of vibration velocities does not exceed substantially the width of resonant line. The figure is drawn for the broadening of resonant line by 1.5 times.

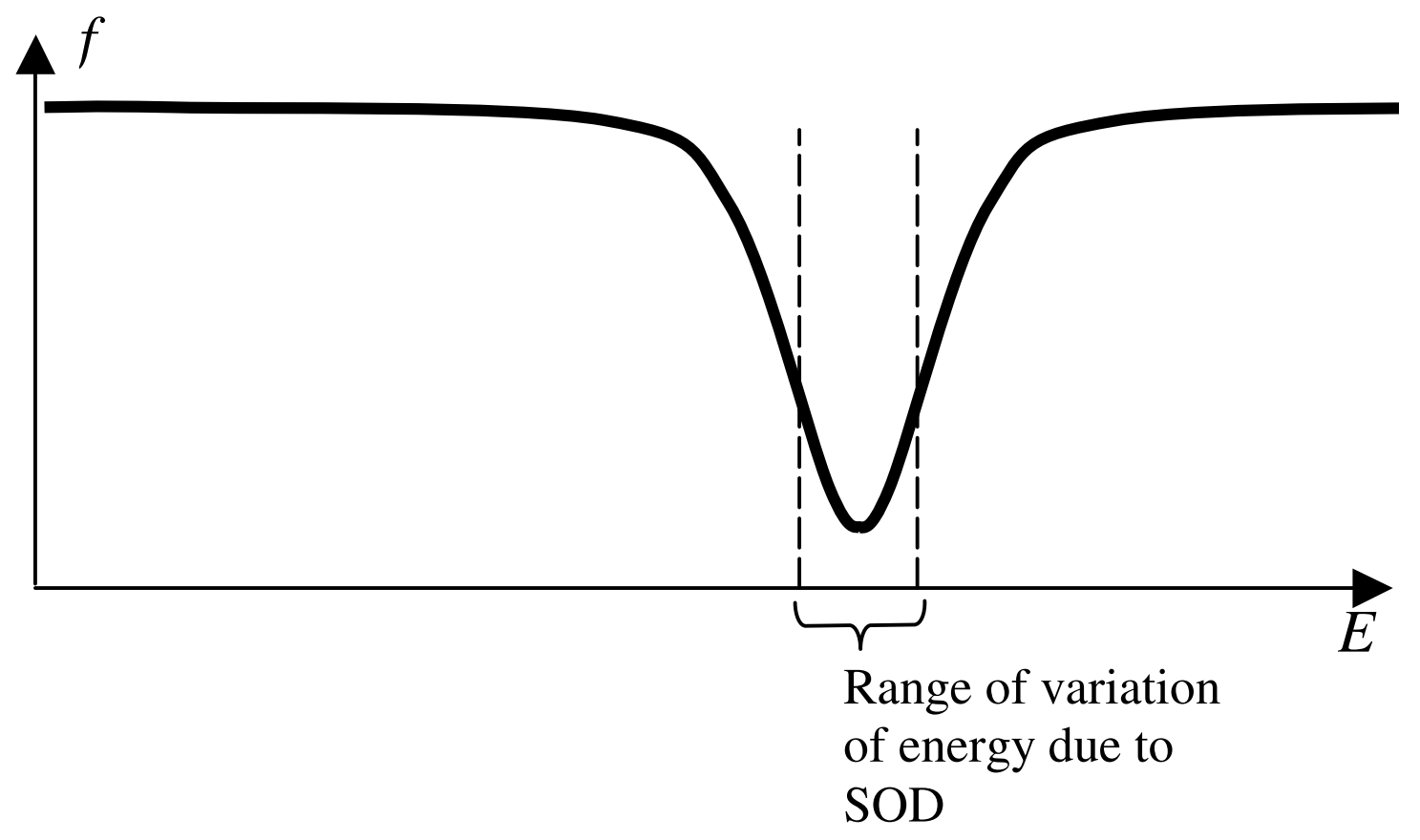

Fig. 2. Resonant line of the absorber 1 and the range of variation of energy due to second order Doppler shift (SOD) for the Mössbauer experiment in rotating system.

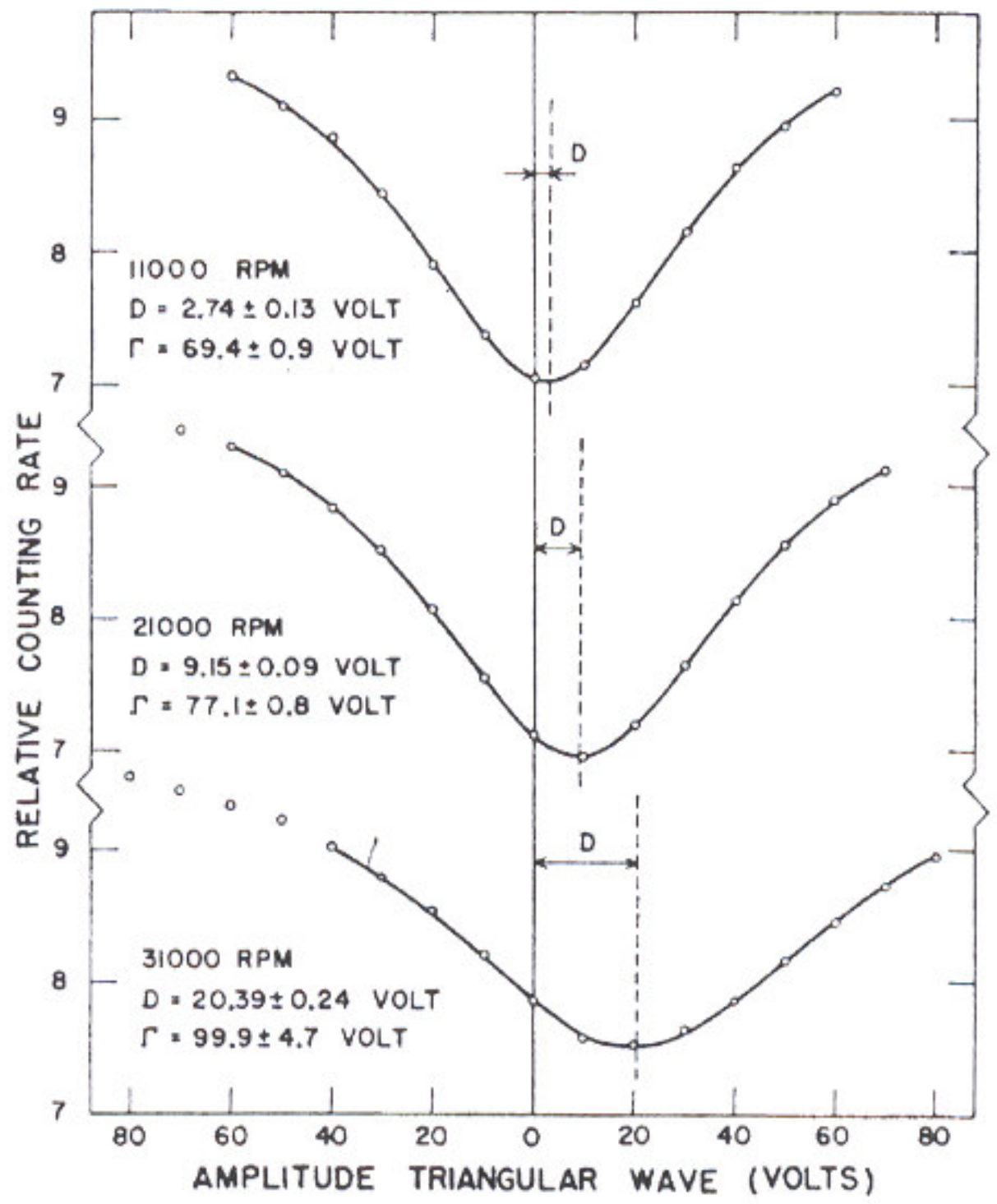

FIG. 3. Typical resonance curves. The amplitude of the triangular wave is proportional to the linear velocity of the source with respect to the absorber. The left (right) side of the plot corresponds to a motion of the source toward (away from) the absorber. The plotted curves are the fitted Lorentz curves normalized to 10 at $v=\infty$. Γ is the full width of the resonance line. With increasing speed of the rotor a considerable broadening of the resonance line was observed. The statistical errors of the points are smaller than the circles.

Fig. 3. Reproduction of Fig. 3 of [1], where a broadening of resonant line with the increase of rotation frequency is clearly seen, without a visible change of its shape. The *y*-axis shows a relative count-rate for three different rotation frequencies, expressed in revolutions per minute (RPM). The axis *x* displays the amplitude triangle wave on the transducer, which is proportional to a linear velocity between the source and absorber in the first order Doppler modulation of the energy of γ-quanta.

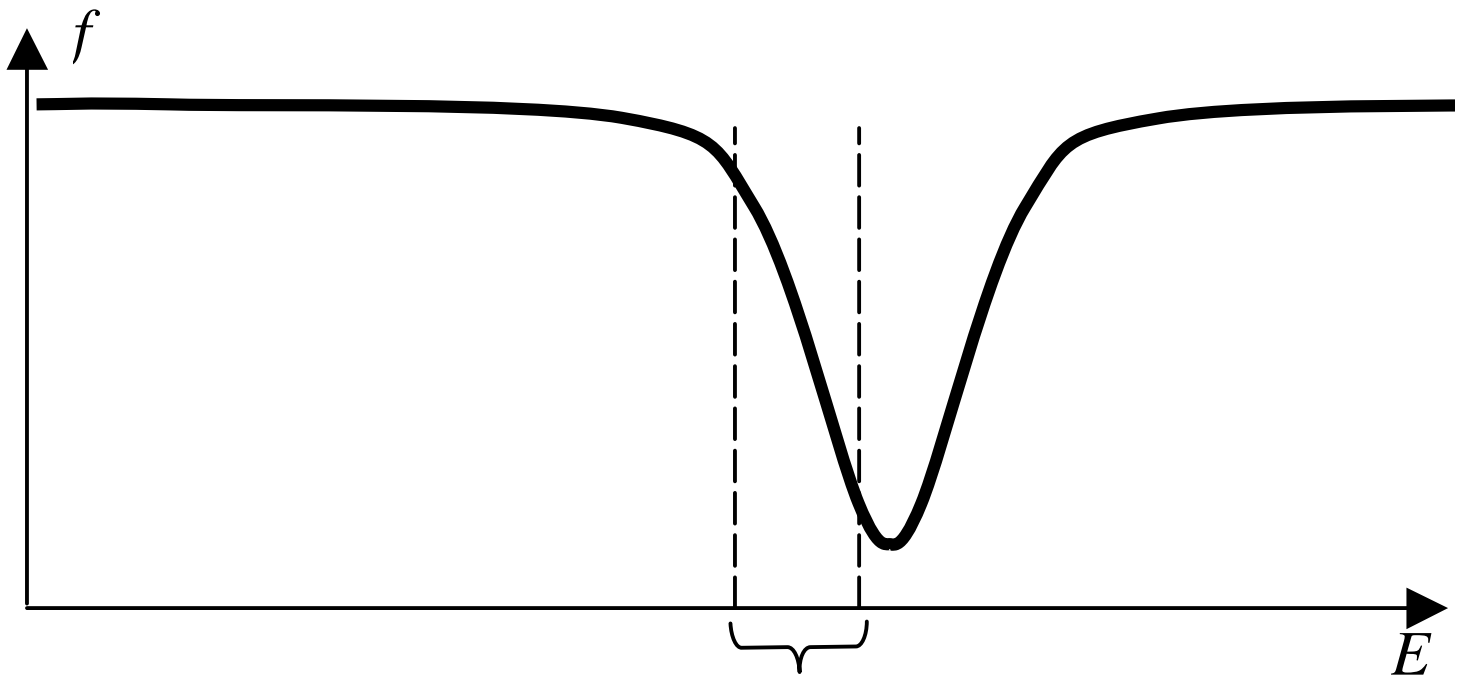

Fig. 4. Resonant line of the absorber 2 and the range of variation of energy due to second order Doppler shift (SOD) for the Mössbauer experiment in rotating system.

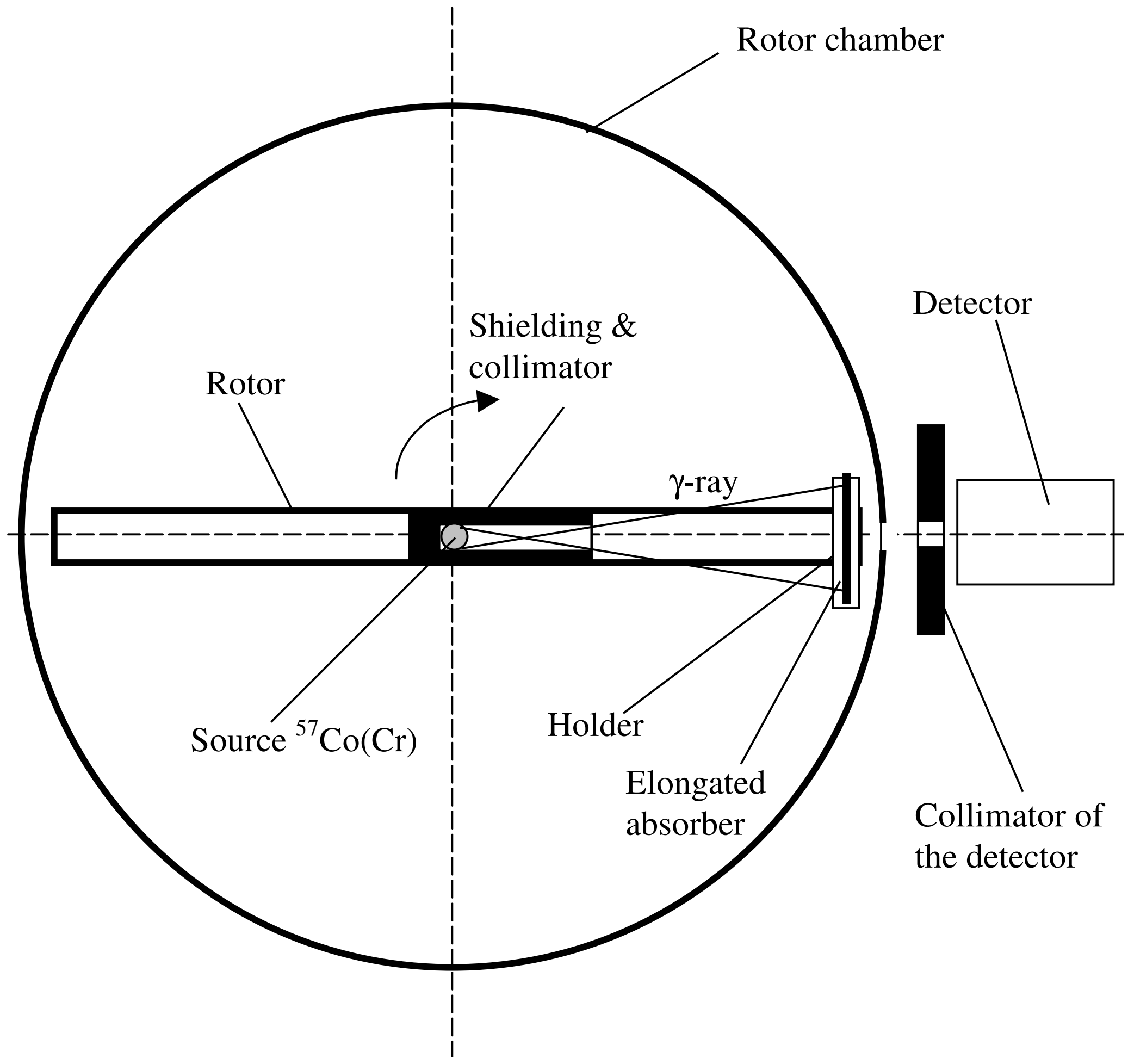

Fig. 5. Schematic of our experiment.

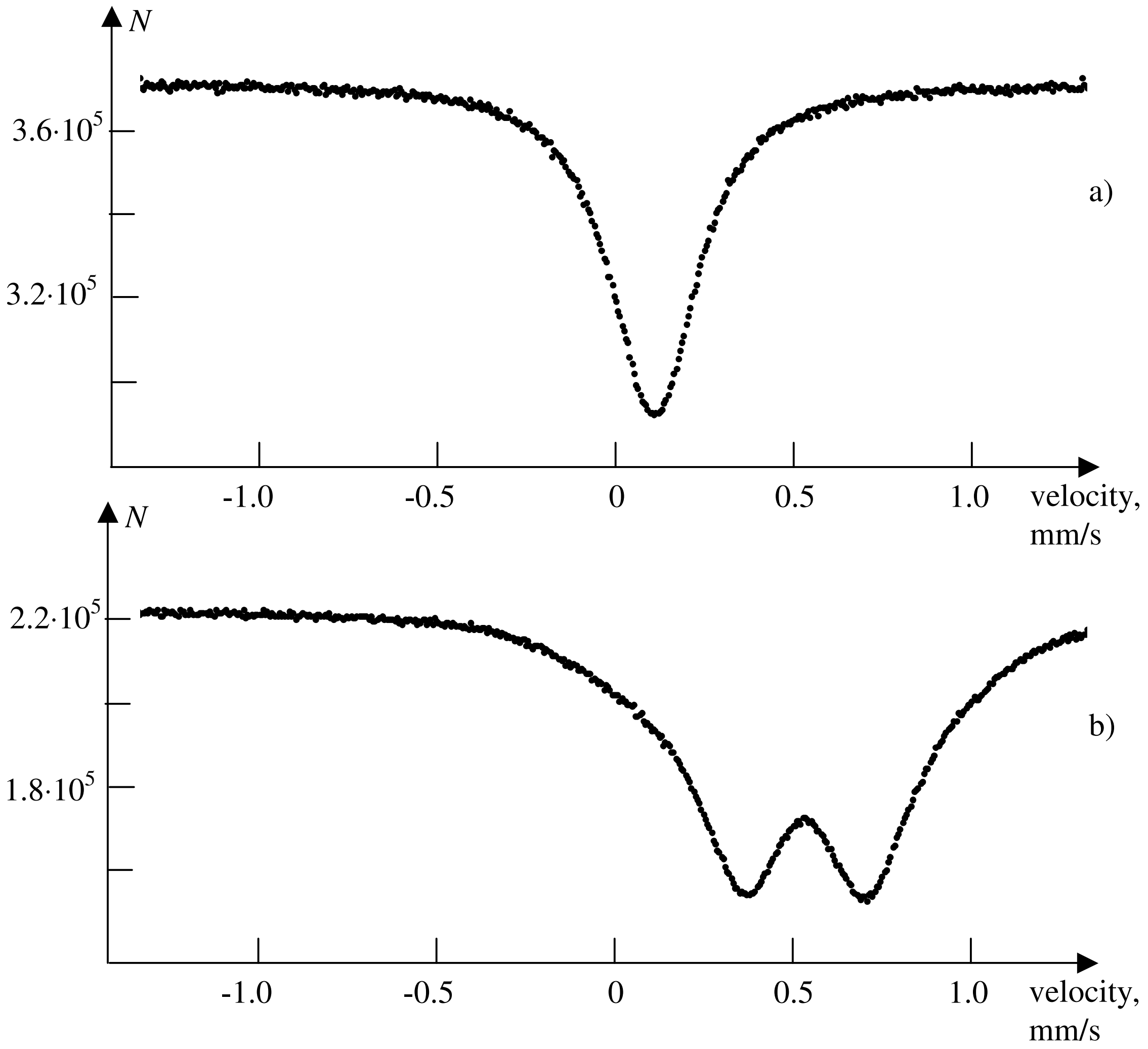

Fig. 6. Mössbauer spectra of $K_4{}^{57}Fe(CN)_6 \times 3H_2O$ (a) and $Li_3{}^{57}Fe2(PO_4)_3$ (b), obtained with the source $^{57}Co(Cr)$.

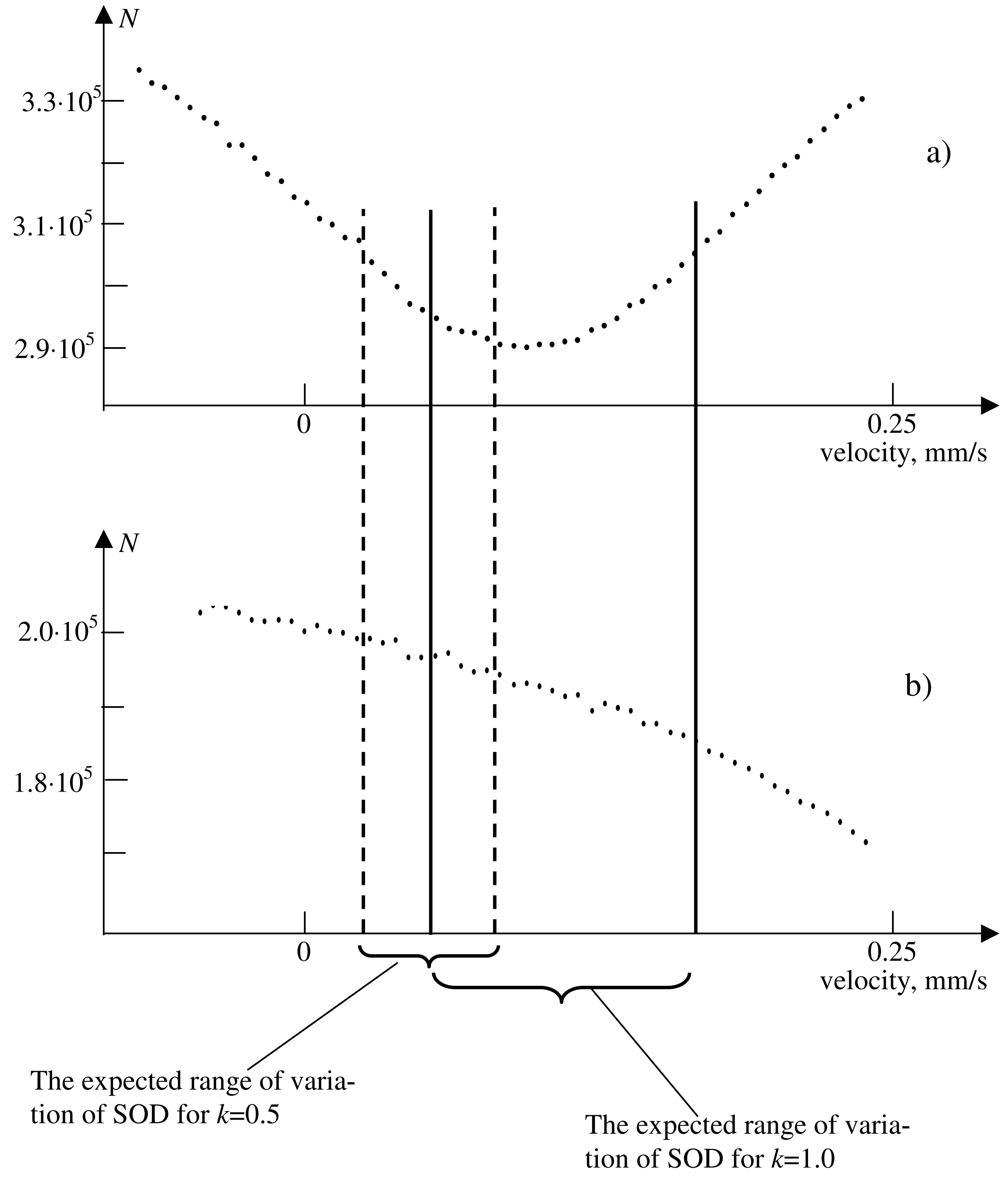

Fig. 7. Expanded relevant fragments of Mössbauer spectra of $K_4{}^{57}Fe(CN)_6 \times 3H_2O$ (a), $Li_3{}^{57}Fe2(PO_4)_3$ (b) and the expected range of variation of SOD in our rotor experiment for two limited hypotheses on *k*.

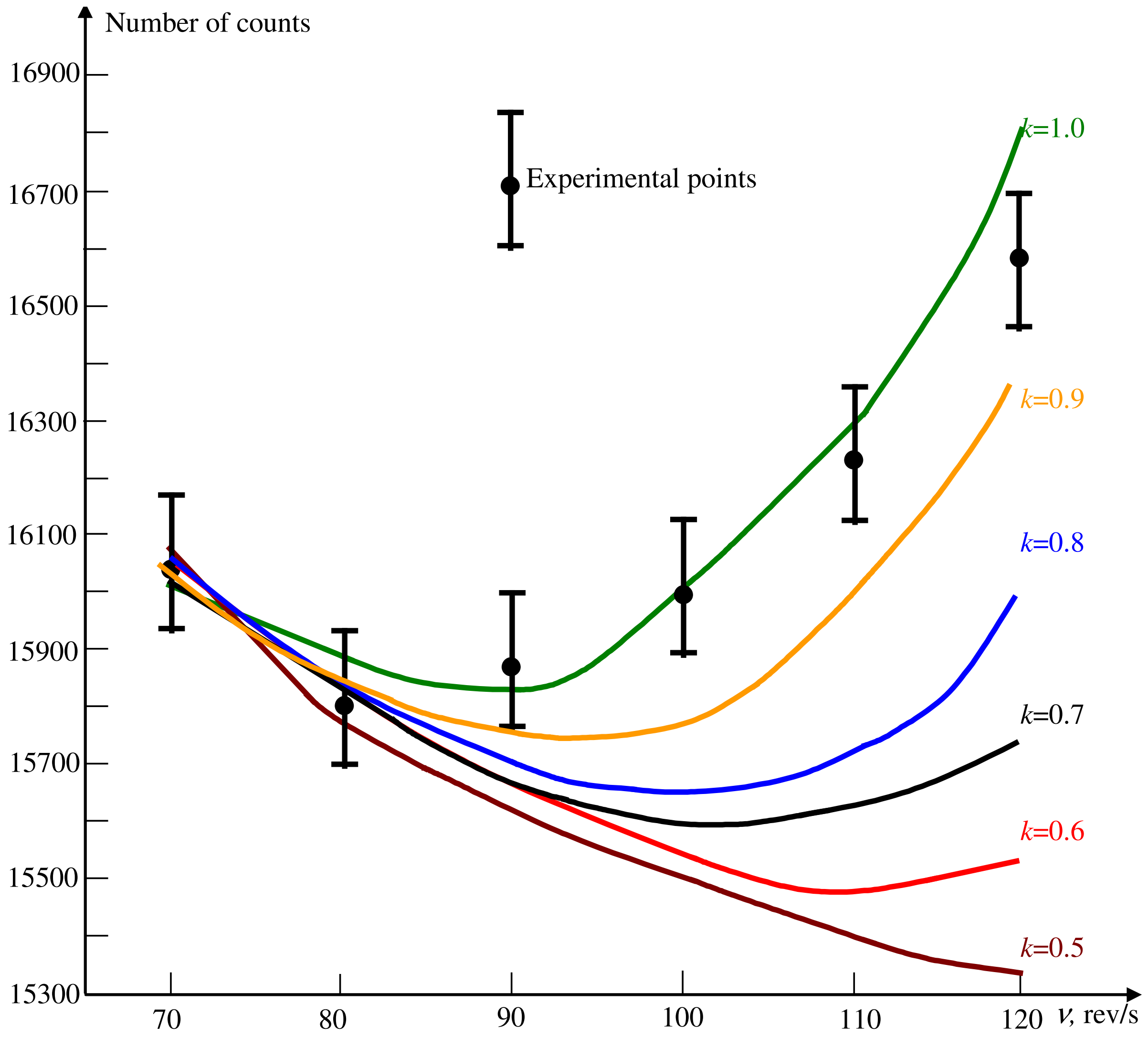

Fig. 8. Experimental data for the absorber 1 in comparison with the curves computed at different $k$ for the idealized (no vibrations) rotor experiment.

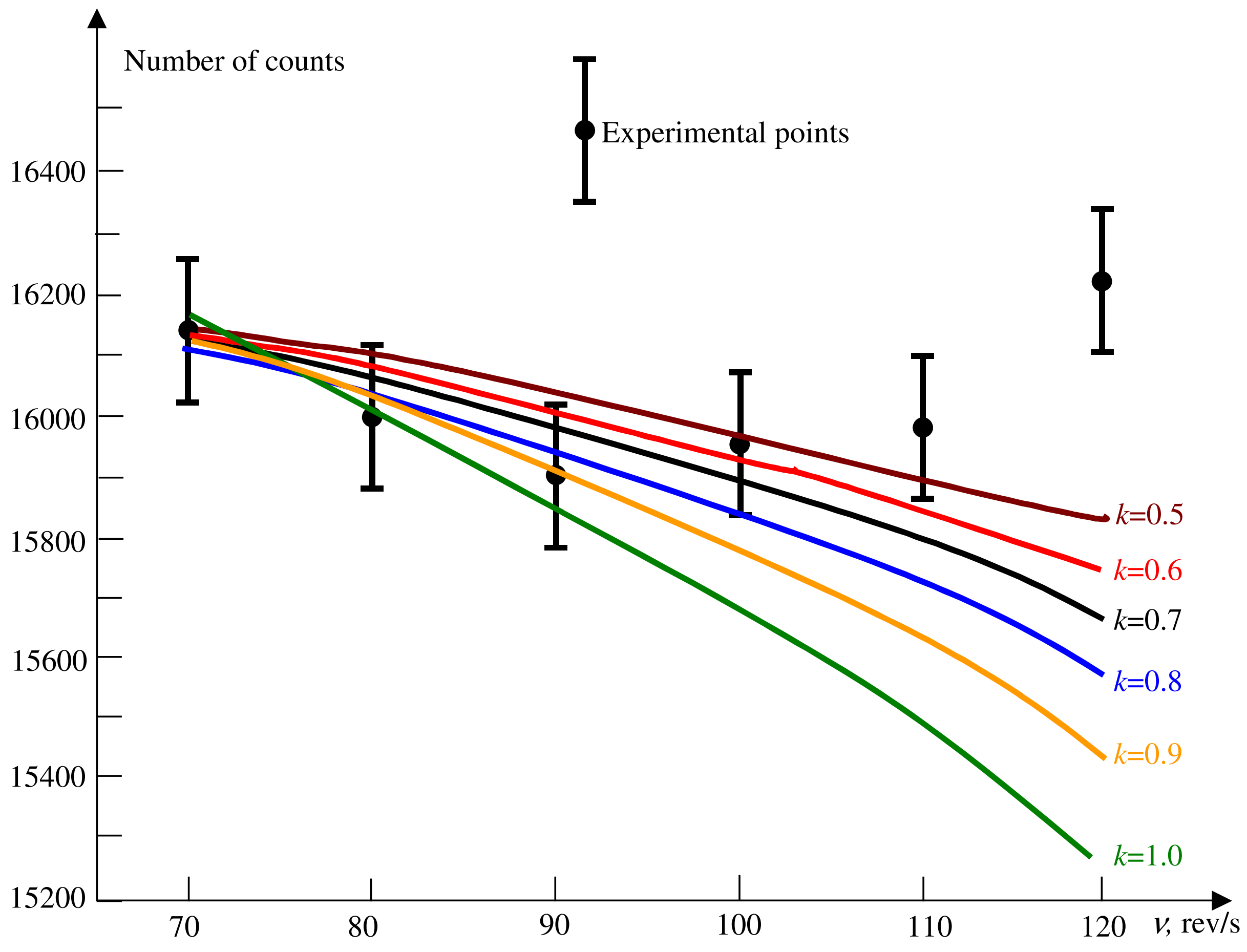

Fig. 9. Experimental data for the absorber 2 in comparison with the curves computed at different $k$ for the idealized (no vibrations) rotor experiment.

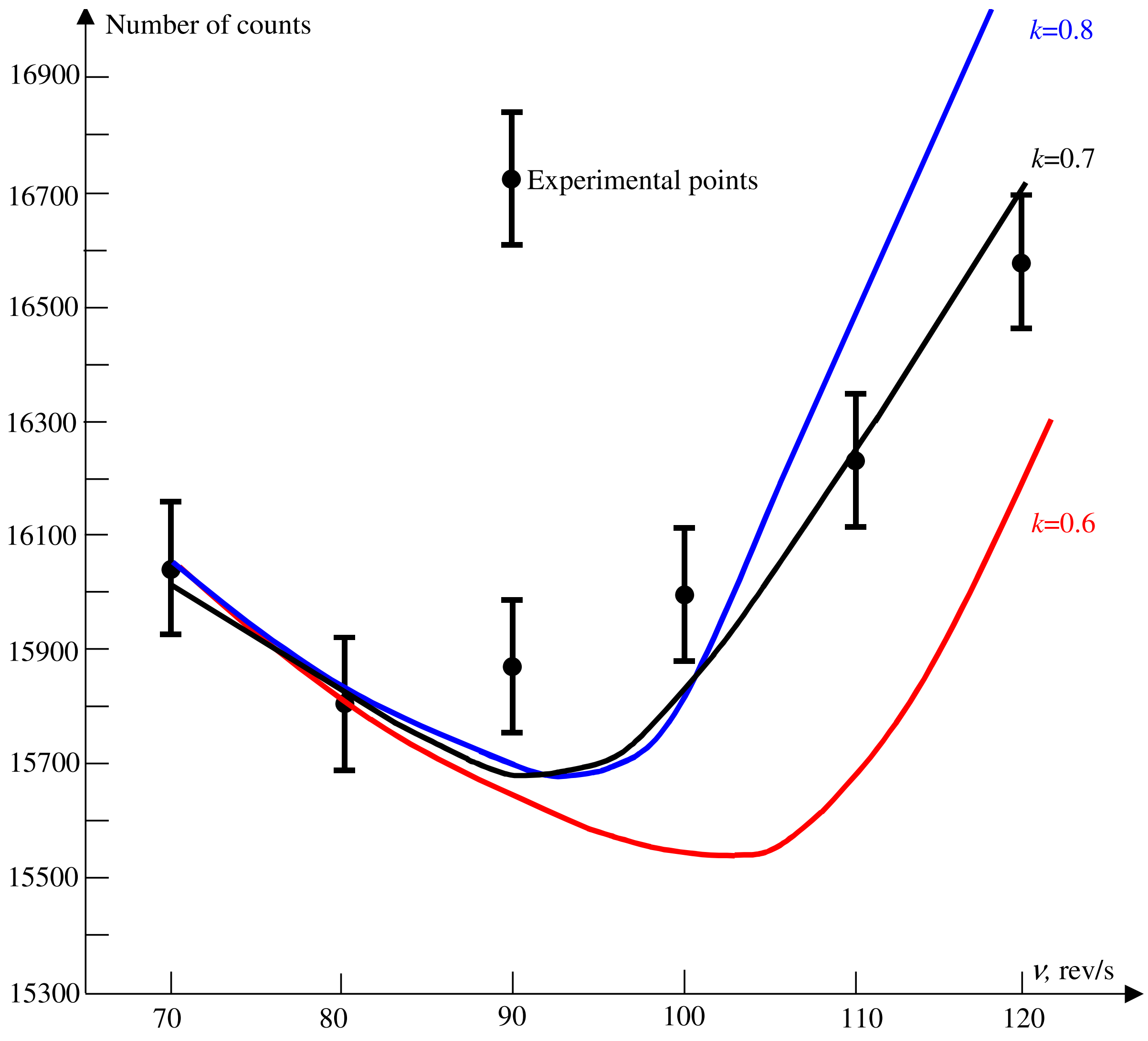

Fig. 10. Experimental data for the absorber 1 and the expected curves re-computed at different $k$ with taking into account the level of vibrations in the rotor system